\documentclass[10pt,conference]{IEEEtran}

\usepackage{setspace}  
\usepackage{anyfontsize} 
\usepackage{pifont}
\usepackage{bbding}
\usepackage{amsfonts}
\usepackage[cmex10]{amsmath}
\usepackage[ruled,vlined,linesnumbered]{algorithm2e}
\usepackage{graphicx}
\usepackage{subfigure}
\usepackage{caption}
\usepackage{stfloats}
\usepackage{float}
\usepackage{xcolor}
\usepackage{booktabs}
\usepackage{tabularx}
\usepackage{amsthm}
\usepackage{xcolor}
\usepackage{setspace}
\usepackage{indentfirst}
\usepackage{amssymb}
\usepackage{setspace}
\usepackage{cite}
\usepackage[normalem]{ulem} 
\usepackage{graphicx}
\usepackage{amsmath, upgreek, booktabs, tabularx, colortbl}
\usepackage{bm}
\usepackage{cases}
\usepackage{cuted}
\usepackage{algorithmicx}
\usepackage{algpseudocode}

\newcommand{\equ}[1]{\begin{equation}#1\end{equation}}

\makeatletter

\newcommand{\Rmnum}[1]{\expandafter\@slowromancap\romannumeral #1@}
\makeatother
\graphicspath{{./Figures/}}

\IEEEoverridecommandlockouts
\usepackage{cite}
\usepackage{amsmath,amssymb,amsfonts}
\usepackage{graphicx}
\usepackage{textcomp}
\usepackage{xcolor}
\def\BibTeX{{\rm B\kern-.05em{\sc i\kern-.025em b}\kern-.08em
		T\kern-.1667em\lower.7ex\hbox{E}\kern-.125emX}}

\begin{document}
	
	\title{Digital Twin-Driven Secure Access Strategy for SAGIN-Enabled IoT Networks}
	
	\author{
		\IEEEauthorblockN{
			Hui Liang\IEEEauthorrefmark{1},
			Zhihui Wu\IEEEauthorrefmark{1},
			Runqi Yuan\IEEEauthorrefmark{1},
			Guobin Zhang\IEEEauthorrefmark{2},
			Yanfeng Zhang\IEEEauthorrefmark{1}
			Jinkai Zheng\IEEEauthorrefmark{1},
			and
			Tom H. Luan\IEEEauthorrefmark{3}
		    }
		\IEEEauthorblockA{\IEEEauthorrefmark{1}School of Electrical Engineering \& Intelligentization,
			Dongguan University of Technology, Guangdong, China}
		\IEEEauthorblockA{\IEEEauthorrefmark{2}Low Altitude Equipment and Intelligent Control Institute, Guangzhou Maritime University, Guangdong, China}
		\IEEEauthorblockA{\IEEEauthorrefmark{3}School of Cyber Science and Engineering, Xi'an Jiaotong University, Shanxi, China}
		\IEEEauthorblockA{Email: \{huiliang, zhihui\_wu, runqiyuan\}@dgut.edu.cn, guobinzh@163.com,  
		yfzhang@ieee.org, \\
		jkzheng@stu.xidian.edu.cn, tom.luan@xjtu.edu.cn
		}
		\thanks{
			This work was supported by Key Area Research and Development Program of Guangdong Province under grant No. 2020B0101110003, in part by General Higher Education Key Area Special Project of Guangdong Province under grant No. 2024ZDZX1045, in part by Dongguan Strategic Scientist Teams Project under Grant No. 20231900700022, and in part by the Guangdong Basic and Applied Basic Research Foundation (2023A1515140003). \emph{(Corresponding author: Hui Liang.)}
		
		}
	}
	
	\maketitle
	
	\begin{abstract}
	In space-air-ground integrated networks (SAGIN)-enabled IoT networks, secure access has become a significant challenge due to the increasing risks of eavesdropping attacks. To address these threats to data confidentiality, this paper proposes a Digital Twin (DT)-driven secure access strategy. The strategy leverages a virtual replica of the physical SAGIN environment within the DT framework to continuously assess dynamic eavesdropping risks by quantifying secrecy capacity. Operating within this DT framework, an evolutionary game model dynamically balances the DT-updated secrecy capacity against queuing delay, steering IoT devices toward more secure and efficient access decisions. Furthermore, a novel distributed algorithm, integral to the DT operation, is developed to obtain the equilibrium access strategy for each device in a scalable manner. Simulation results demonstrate that the proposed DT-based approach substantially improves the security of SAGIN-enabled IoT networks. Additionally, it effectively balances system load, prevents overload occurrences, and decreases queuing delay compared to benchmark schemes, thereby comprehensively improving overall network performance.
	\end{abstract}
	
	
	\IEEEpeerreviewmaketitle
	
	\section{Introduction}
	
	\IEEEPARstart{T}{he} The proliferation of IoT devices demands ubiquitous connectivity, driving the integration of space-air-ground communication infrastructures into SAGIN-enabled IoT networks \cite{cui2022space,liu2018space}. The SAGINs promise unprecedented capabilities by synergistically leveraging low Earth orbit satellites for global coverage and low latency, aerial platforms (e.g., Unmanned Aerial Vehicles (UAVs), High Altitude Platform Station (HAPS)) for flexible deployment and line-of-sight advantages, and terrestrial networks for high capacity and dense connectivity\cite{10829750}. This diverse architecture expands the scope of applications requiring secure and reliable data transmission, especially in areas lacking traditional infrastructure, such as disaster zones, maritime environments, remote industrial sites, and underserved regions \cite{tang2021deep}. A key component of SAGINs is edge computing, which offloads processing tasks from resource-limited IoT devices to distributed edge service nodes (ESNs) located on satellites, aerial platforms, or ground stations, aiming to reduce latency and improve performance \cite{tang2021deep,zhang2025block,jia2024service}.

However, SAGINs face significant security challenges due to their heterogeneous and dynamic nature, such as high-speed mobility, limited resources, intermittent connectivity, and varying node capabilities \cite{10118696}. These issues threaten the reliability of edge computing offloading, mainly due to reliance on open wireless channels across space-air-ground segments \cite{zhang2025basis,kang2024survey}. Sensitive data—like industrial commands or health records—is at risk from multiple threats: data leakage that compromises systems, interference or spoofing that disrupts operations, and eavesdropping by malicious nodes (e.g., compromised satellites, UAVs, or ground stations), which poses the greatest threat to confidentiality \cite{kang2024survey}.  

These risks are severe in mission-critical applications—such as government operations, industrial automation, and emergency response—where data breaches could endanger national security or public safety \cite{10118696}. Therefore, proactive risk assessment and strong access control are essential to reduce eavesdropping risks and ensure secure and reliable ESN selection.

	To address these challenges in SAGIN-enabled IoT networks, we propose an efficient service selection framework tailored for scenarios involving space-based eavesdropping satellites. However, the large scale, heterogeneity, and rapid dynamics of SAGINs impose significant constraints. To accurately capture and predict these dynamic risks, we incorporate the concept of Digital Twin (DT)\cite{grieves2014digital,10901598}. A DT creates a virtual replica of the physical SAGIN environment, encompassing satellite states, device distributions and channel conditions. This virtual representation facilitates continuous, real-time assessment of security threats, such as eavesdropping risks. Given the impracticality of centralized access control in large-scale IoT networks with rapidly changing topologies and node conditions, we adopt evolutionary game theory \cite{tanimoto2015fundamentals} to optimize confidentiality and delay under these constraints jointly. It formulates how IoT devices adjust their strategies to reach a balanced state. This method enables distributed devices to dynamically adapt their strategy based on local utility feedback, achieving a balanced state without the need for global coordination. Moreover, it can effectively balance security and latency in resource-constrained scenarios and provides a scalable and self-organizing solution for dynamic edge computing in heterogeneous SAGIN environments.

	The main contributions of this work are as follows:
	
	\begin{enumerate}
		\item A novel Digital Twin-based secure access selection strategy for SAGIN-enabled IoT networks is proposed. The DT framework incorporates security factors by quantifying eavesdropping risks, enabling secure and efficient edge computing access decisions.
		
		\item Within the DT, an evolutionary game-theoretic service selection strategy is designed to jointly balance secrecy capacity (derived from the risk model) and queuing delay (derived from the resource model), driving the network towards a stable equilibrium state. 
		
		\item Comprehensive verification of the proposed solutions through extensive simulation experiments demonstrates their effectiveness in enhancing system security and performance.
		
	\end{enumerate}

	The remainder of this paper is structured as follows.  Section II describes the system model for the SAGIN-enabled IoT networks framework.  Section III details the proposed evolutionary game-theoretic framework for ensuring secure and efficient service selection.  Section IV designs a distributed selection algorithm.  Extensive simulation results are shown in Section V.  Finally, Section VI concludes this work and gives future research directions.

	\section{System Model}
	\label{Sec:SystemModel}
	
	
	\begin{figure}[t]
		\centering
		\includegraphics[scale=0.54,trim=0 0 0 0]{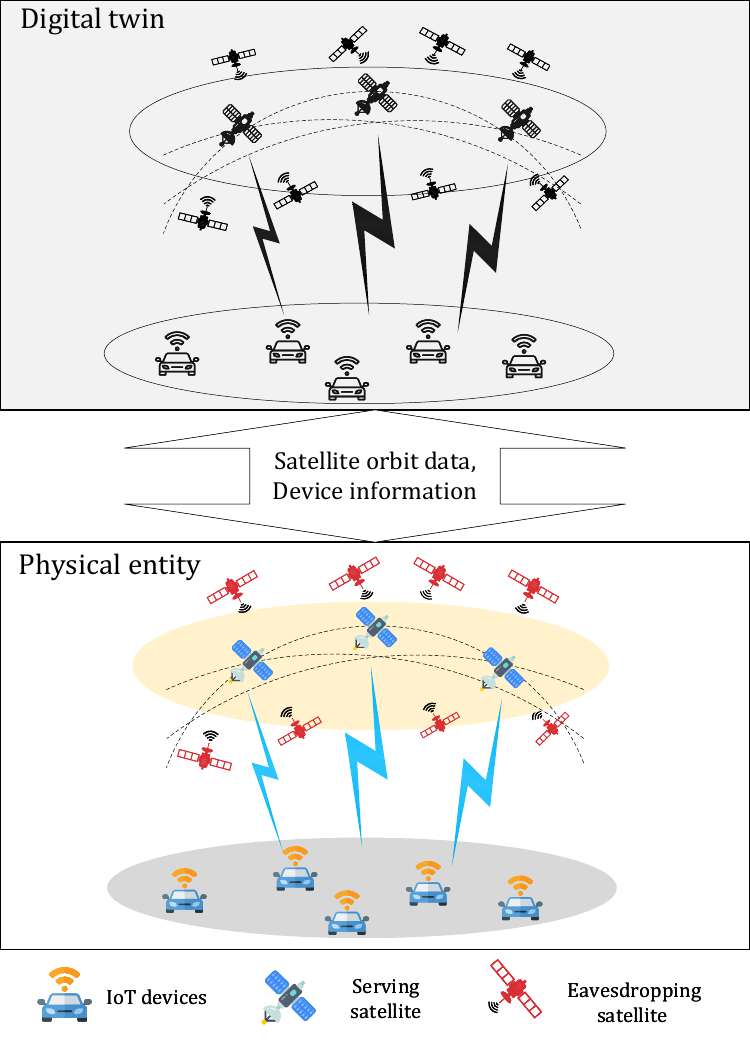}
		\caption{System model of Digital Twin-assisted service selection in the SAGIN-enabled IoT networks }
		\label{Fig:SystemModel}
	\end{figure}

	In this section, the basic construction of the SAGIN-enabled IoT network is introduced, including the space layer, the ground layer, and their integrated network model.
	
	As shown in Fig.~\ref{Fig:SystemModel}, the SAGIN-enabled IoT network is augmented by a Digital Twin layer to provide computing support for resource-constrained terrestrial IoT devices in remote areas without terrestrial network coverage.

	The physical domain includes $M$ service satellites, indexed by $i = 1, 2, \dots, M$, which offer edge computing services to $N$ IoT devices, indexed by $n = 1, 2, \dots, N$. These IoT devices are distributed across the coverage areas of multiple satellites and offload their computing tasks to them. The system also faces security threats from an unknown number of eavesdropping satellites, which are typically positioned in specific orbits and operate near service satellites to intercept transmitted information.  
	
	The Digital Twin maintains a synchronized virtual representation of these entities and their interactions. It continuously receives data (e.g., satellite ephemeris, estimated channel states, device connection requests, satellite load reports) from the physical domain to update its state. Key metrics such as secrecy capacity $C_i^S(t)$ and queuing delay $D_i(t)$ are computed and dynamically updated in this virtual environment based on the current state.
	
	Given this complex network topology, each IoT device must dynamically select an optimal serving satellite based on both security considerations and queuing efficiency. At each selection round, a device can only connect to one satellite and is allowed to change its selection in subsequent rounds. Let $x_i(t) \in [0,1]$ denote the proportion of devices selecting satellite $i$ at time $t$.  The total number of devices connected to satellite $i$ at time $t$ is given by $b_i(t) = x_i(t)N$. The network selection strategy is influenced by both service demand and network characteristics, which are assumed to remain stable within a given selection round. By considering factors that are security risks and queuing delays, IoT devices continuously adjust their satellite selection strategy to improve system performance and enhance their security.
	
	\subsection{Secrecy Capacity}
	
	Secrecy capacity is a key metric in satellite communication security, quantifying the difference between the legitimate transmission capacity and the total intercepted capacity of potential eavesdroppers. The spatial distribution of eavesdropping satellites relative to IoT devices has a direct impact on the probability of communication interception. We use the shadow-Rician fading model as the channel between the device and the satellite, while assuming that the eavesdropper satellites are uniformly distributed on a spherical shell with radius \(R + H_e \), where $R$ is the Earth's radius and $H_e$ is the orbit altitude of the eavesdropper satellite. Secrecy capacity is defined as the maximum information rate that a legitimate communication link can safely transmit in the presence of an eavesdropper. It quantifies the maximum transmission rate that a communication system can achieve while ensuring information security. The secrecy capacity $C_i^{\mathrm{S}}(t)$ for serving satellite $i$ at time $t$ is given by:
	
	\begin{equation}
		C_i^{\mathrm{S}}(t) = \left[\log _2\left(1+\gamma_i(t)\right)-C_i^E\right]^{+},
	\end{equation}
	where $\gamma_{i}(t)$ is the SNR of the legitimate channel. $C_i^E$ is the total eavesdropping capacity, representing the maximum information rate that can be intercepted by the eavesdropper. The symbol $[x]^{+}$ ensures that the secrecy capacity is a non-negative value, which means that secure transmission can only be achieved when the legitimate transmission rate exceeds the eavesdropping capacity. The $C_i^E$ can be obtained by following the steps.
	
	First, we define the eavesdropping capacity of a satellite at a specific distance $d_e$ (denoted as $C_i^e$) using the following formula:
	\begin{equation} 
		C_i^e = W \log_2\left(1 + \gamma(d_e)\right),
	\end{equation}
	where \(W\) is the channel bandwidth, and \(\gamma(d_e)\) denotes the SNR at distance \(d_e\). 
	
	Nevertheless, being within the line-of-sight range does not guarantee successful interception. For effective eavesdropping to occur, the eavesdropper must also reside within the effective coverage cone of the service satellite. This region is called the effective eavesdropping area for the target service satellite. It is formally defined as a spherical cap centered on the service satellite. The angular radius of this cap is denoted as $\psi_{\text{eff},i}$, which is derived as follows:
	\begin{equation}
		\psi_{\text{eff},i} = \cos^{-1} \left( \frac{R}{R + H_i} \right).
	\end{equation}
	In Eq.(3),  $H_i$ is the orbital altitude of the service satellite. This relationship highlights that satellites at higher altitudes have wider coverage areas, thereby increasing their vulnerability regions. 
	
	Thus, the probability that a randomly positioned eavesdropper poses a threat to the $i$-th service satellite is given by:
	\begin{equation}
		p_i^S = \frac{S_{\Psi_{\text{eff},i}}}{S_c} = \frac{1 - \cos \psi_{\text{eff},i}}{2},
	\end{equation}
	where $\psi_{\text{eff},i} = \cos^{-1} \left( \frac{R}{R + H_i} \right)$ is the vulnerable surface area,and $S_c = 4\pi (R + H_i)^2$ is the total surface area of the sphere at the orbital altitude of the service satellite.

	After this, all we need to know is how many eavesdropping satellites exist within this effective range, and then we can calculate the total eavesdropping capacity. To describe the number of eavesdropping satellites using a probability distribution, we first define the combined probability $\xi_i$ that a single eavesdropping satellite is both within the distance interval $[d_e, d_e + \Delta d_e]$ and within the effective eavesdropping region of the $i$-th serving satellite as:
	\begin{equation}
		\xi_i = f_d(d_e) \cdot \Delta d_e \cdot p_i^S. 
	\end{equation}
	
	In Eq.(5), $f_d(d_e)$ is the probability density function of the distance $d_e$ obtained the probabilistic geometric model through as follows:
	
	\begin{equation}
		f_d(d_e) = \frac{d_e}{2R(R + H_e)}.
	\end{equation}
	
	Therefore, assuming that there are $K$ eavesdropping satellites evenly distributed around the service space, we have derived the following expression regarding the expected number of eavesdropping satellites:
	\begin{equation}
		\mathbb{E}[\Xi(d_e)] =
		\begin{cases}
			0, & d_e \leq H_e, \\
			K \cdot \frac{d_e}{2R(R + H_e)} \cdot \Delta d_e \cdot p_i^S, & H_e < d_e \leq D_{\text{max}}, \\
			0, & d_e > D_{\text{max}}.
		\end{cases}
	\end{equation}
	
	Finally, the total eavesdropping capacity $C_i^E$ under this configuration is given by:
	\begin{equation}
		\begin{split}
			C_i^E &=\int_{H_e}^{D_{\text{max}}} \mathbb{E}[\Xi(d_e)] \cdot C_i^e \, \mathrm{d}d_e \\
			&= \frac{K W p_i^S}{2R(R + H_e)} \int_{H_e}^{D_{\text{max}}} d_e \log_2\left(1 + \gamma(d_e)\right) \, \mathrm{d}d_e.
		\end{split}
	\end{equation}

	
	\subsection{Queuing Delay}
	
	In satellite networks, the queuing delay experienced by IoT devices refers to the computing waiting time in the satellite queue. This delay is particularly significant when multiple IoT devices compete for limited resources, as it directly impacts the overall system performance and efficiency.
	
	To model this queuing delay, we use the \( M/M/1 \) queuing model \cite{giambene2005queuing}. The resulting queuing delay \(D_i(t)\) can then be formulated as:
	
	\equ{
		\label{Equ-QueuingDelay}
		D_i(t) = \frac{1}{\mu_i - \lambda_i(t)},
	}
	where \( \lambda_i(t) \) is the arrival rate of device tasks to the queue of satellite $i$ at time $t$, \( \mu_i \) is the service rate of the satellite $i$, representing the rate at which tasks are processed, and \( D_i(t) \) is the queuing delay for satellite \( i \) at time \( t \).
	\noindent As the system approaches its peak processing limit (\( \lambda_i(t) \to \mu_i \)), delays become more evident, adversely affecting performance.

	\section{Problem Formulation}
	\label{Sec:Problem Formulation}
	
	In this section, in order to address the challenges brought about by the dynamic changes in the SAGIN-enabled IoT networks environment and to achieve the equilibrium of secrecy capacity and queuing delay, we model the service selection problem as an evolutionary game\cite{tanimoto2015fundamentals}.
	
	\subsection{Evolutionary Game Formulation}
	Evolutionary game theory is a mathematical framework that models strategic interactions within populations of agents, where strategies evolve over time based on their relative success (fitness or payoff) rather than assuming perfect rationality\cite{zhang2022cybertwin,fan2022network}.
	
	The evolutionary game of service selection in the SAGIN-enabled IoT network system can be formulated as follows.
	
	\begin{itemize}
		\item \emph{Players}: A player is defined as an IoT device that has more than one selection of serving satellites in an area. In this scenario, we only refer to devices with multiple accessible serving satellites as players. 
		
		\item \emph{Strategy}: The strategy of each player is the decision of service satellite selection. In the service area, the set of strategies can be denoted by $s\in\mathcal{S}=\{1,2,…,M\}$.
		
		\item \emph{Population}: The population is formed by the set of players in the service area and the number of populations is assumed to be finite. The evolutionary game is played among populations rather than players.
		
		\item \emph{Population Share}: The population share is referred to as the ratio of the number of devices that choose a certain strategy to the total number of devices in the system. Thus, the number of devices who choose strategy $i$ can be expressed as $s_i(t) = x_i(t)N$, where $x_i(t)$ is the proportion of devices selecting the $i$-th service satellite at time $t$.

		\item \emph{Payoff}: The payoff of each player is the utility of each device, which is influenced by secrecy capacity and queuing delay. We use the linear combination of the above two factors as the utility function.
	\end{itemize}
	
	\subsection{Utility Function}
	
	The utility function is used to measure the device's satisfaction under different strategies. The higher the value of the utility function, the higher the device's satisfaction with the corresponding strategy selection. In this paper, we use \textit{secrecy capacity} $C^S_i (t)$ and \textit{queuing delay} $D_i (t)$ to define the utility $\pi_i(t)$ of an IoT device when selecting satellite $i$ at time $t$ as follows:
	
	\begin{equation}
		\label{Equ:Utility}
		\pi_i (t) = \alpha \frac{C^S_i (t)}{x_i (t) N} - \beta D_i (t),
	\end{equation}
	
	\noindent where the first term, $\alpha\frac{C^{S}_{i}(t)}{x_{i}(t)N}$, represents the normalized secrecy capacity per device (which diminishes as more devices select the same satellite), and the second term, $-\beta D_{i}(t)$, penalizes the queuing delay. The weighting factors $\alpha$ and $\beta$ balance the relative importance of security and latency in the access strategy.

	\subsection{Average Utility}
	
	The average utility serves as a metric to evaluate the overall system efficiency.  It is calculated as the weighted sum of each population's utility, with weights based on device proportions.

	The average utility of the system at time $t$, denoted as $\bar{\pi}(t)$, is given by:
	
	\begin{equation}
		\label{Equ-ExceptedUtility}
		\bar{\pi}(t) = \sum_{i=1}^M x_i(t) \pi_i(t), 
	\end{equation}
	where  $\pi_i (t)$ represents the utility of the IoT device when satellite $i$ is selected at time $t$ , while $x_i(t)$ is the proportion of devices that choose service satellite $i$ at time $t$. Higher average utility $\bar{\pi}(t)$ indicates improved system efficiency, enhanced security, and reduced latency. When the utility obtained by the device is lower than the average utility of the system, the device will adjust its service selection strategy to provide users with more optimal choices. 
	\subsection{Replicator Dynamics for Evolutionary Game}
	
	In the SAGIN-enabled IoT network systems, access selection strategies for IoT devices are dynamically optimized through the application of dynamic replication within evolutionary game theory. Devices adjust their access strategy based on real-time utility evaluations and are more likely to select serving satellites with higher utility values. Replicator dynamics describes the temporal variation in the proportion of devices selecting a specific satellite. The replicator dynamics for the population share of devices selecting satellite relay $i$ at time $t$, denoted as $\dot{x}_i(t)$, is expressed as:
	
	\equ{\label{Equ-ReplicatorDynamics}
		\dot{x}_i(t) = \sigma x_i(t) \left( \pi_i(t) - \bar{\pi}(t) \right).
	}
	
	The replicator dynamics ensure that its proportion in the population will increase when a strategy’s utility exceeds the system’s average utility; conversely, its proportion will decrease if a strategy’s utility is below the average. This feedback mechanism drives the system toward an evolutionary equilibrium, where all strategies either attain equal utility or become deselected.
	
	\subsection{Evolutionary Equilibrium}
	The replicator dynamics in Eq.(12) drive the strategic adaptation of IoT devices toward a stable configuration. This dynamic process converges when the population distribution reaches an evolutionary equilibrium, defined as a state where no unilateral strategy change can improve individual utility. Mathematically, this equilibrium satisfies: 
	
	\equ{\label{ESS}
		\pi_i(t) = \bar{\pi}(t), \quad \forall i \in \{1, 2, \dots, M\}.
	}
	
	In this state, the system achieves load balancing and efficient resource utilization, ensuring that the device complies with security requirements while minimizing queuing delays. 
	
	\section{Distributed Iterative Service Selection Algorithm}
	
	The service selection problem for IoT devices is formulated and addressed using an evolutionary game approach within the Digital Twin (DT) framework. The DT provides real-time secrecy capacity $C_i^S(t)$ and queuing delay $D_i(t)$ for each satellite $i$, which are used as key inputs in the device utility function (Eq. 10). The replicator dynamics (Eq. 12) update the population state in the DT. The distributed selection algorithm (Algorithm 1), either executed or supported by the DT, guides device strategies toward evolutionary equilibrium based on the updated virtual state. This algorithm uses the replicator dynamics to enable IoT devices to adapt their strategies in response to eavesdropping threats, balancing security and queuing delay.
	
	This distributed algorithm first allocates the initial device proportions among the satellites and then enters the iterative cycle stage. During this stage, the satellites first broadcast their instantaneous number of device connections to disseminate load information. After receiving this global state information, each IoT device considers secrecy capacity and queuing delay as key components and calculates its utility $\pi_i(t)$ for all accessible satellites. Subsequently, the devices calculate the average utility $\bar{\pi}(t)$ of the entire system, thereby establishing an evolving benchmark. 
	
	This utility assessment will lead to strategic adjustments. At the current moment $t$, devices that choose the utility of satellite $i$ to be lower than the average utility (i.e.,$\pi_i(t) < \bar{\pi}(t)$) will probabilistically migrate to other satellites, while those with utilities exceeding the average (i.e.,$\pi_i(t) > \bar{\pi}(t)$) will maintain their current association. This dynamic adjustment process is controlled by the replicator dynamic mechanism in Eq.(12), which enables the device strategy to continuously optimize and ultimately maximize the overall utility of the system. As the IoT devices adjust their selections based on real-time feedback, the population proportions stabilize, and the system reaches an evolutionary stable state where no device has an incentive to deviate. This process ensures optimal load balancing across satellites, reduces queuing delays, and enhances system security by dynamically adapting to changes in network conditions.

	\begin{algorithm}[t]
		\label{Sec:EvolutionaryGameAlgm}

		\caption{Seeking Equilibrium in Serving Satellite Selection Game}
		\label{algo:AlgorithmProcedure}
		\SetAlgoLined
		\SetKwInOut{Require}{Require}\SetKwInOut{Ensure}{Ensure}\SetKwInOut{Output}{Output}
		\SetKwData{Left}{left}\SetKwData{This}{this}\SetKwData{Up}{up}
		\Require{Initial population proportions $\mathbf{x}(0)$ and system model parameters}
		\Ensure{}
			\For{timeslot $t=1,..., T$}{
				Satellites broadcast the number information of devices connected to them \\
				\For {each device $n \in \mathcal{N}$}{
					\For {strategy $s \in \mathcal{S}$}{
						Device $n$ calculates its utility $\pi_i(t)$ and broadcast its utility information \\
						Device $n$ receives the number information of devices connected to each satellite, then calculate the average utility $\bar{\pi}(t)$ \\
						\eIf{$\pi_i(t) < \bar{\pi}(t)$}
						{Device $n$  select a different satellite with a fixed probability} 
						{
							Device $n$ keep the selection unchanged \\
						}
					}
				}
			}
			\Output{Equilibrium population proportions}
	\end{algorithm}
	
	The computational complexity of Algorithm 1 arises from its iterative process, which involves multiple loops across time slots, the number of devices, and the number of strategies. Key advantages of the proposed algorithm include its scalability and flexibility. By enabling devices to self-organize and adapt their strategies in response to real-time system feedback, the algorithm effectively mitigates the challenges posed by eavesdropping threats and resource constraints in SAGINs.

	\section{Simulation Results}
	\label{Sec:Experimental Results}

  To evaluate the proposed serving satellite selection strategy for the SAGIN-enabled IoT network, we constructed a simulation environment to reflect the core characteristics of SAGINs comprehensively. The simulation scenario is set in a remote area without ground network coverage, including four serving satellites with edge computing capabilities and several eavesdropping satellites with malicious attacks. The IoT devices are randomly distributed within the coverage of the serving satellites, with the number of devices $N$ ranging from 100 to 1000, aiming to reproduce diverse actual communication scenarios. The serving satellites are evenly distributed on an orbit at 300 kilometers height, with initial phases of 0°, 90°, 180°, and 270°, respectively, while the eavesdropping satellites are located on an orbit at 600 kilometers height, with an initial phase of 0°.
	
	\subsection{Average Utility Comparison}
	
	\begin{figure*}[t] 
		\centering
		\begin{minipage}{0.3\textwidth}
			\centering
			\includegraphics[width=\linewidth]{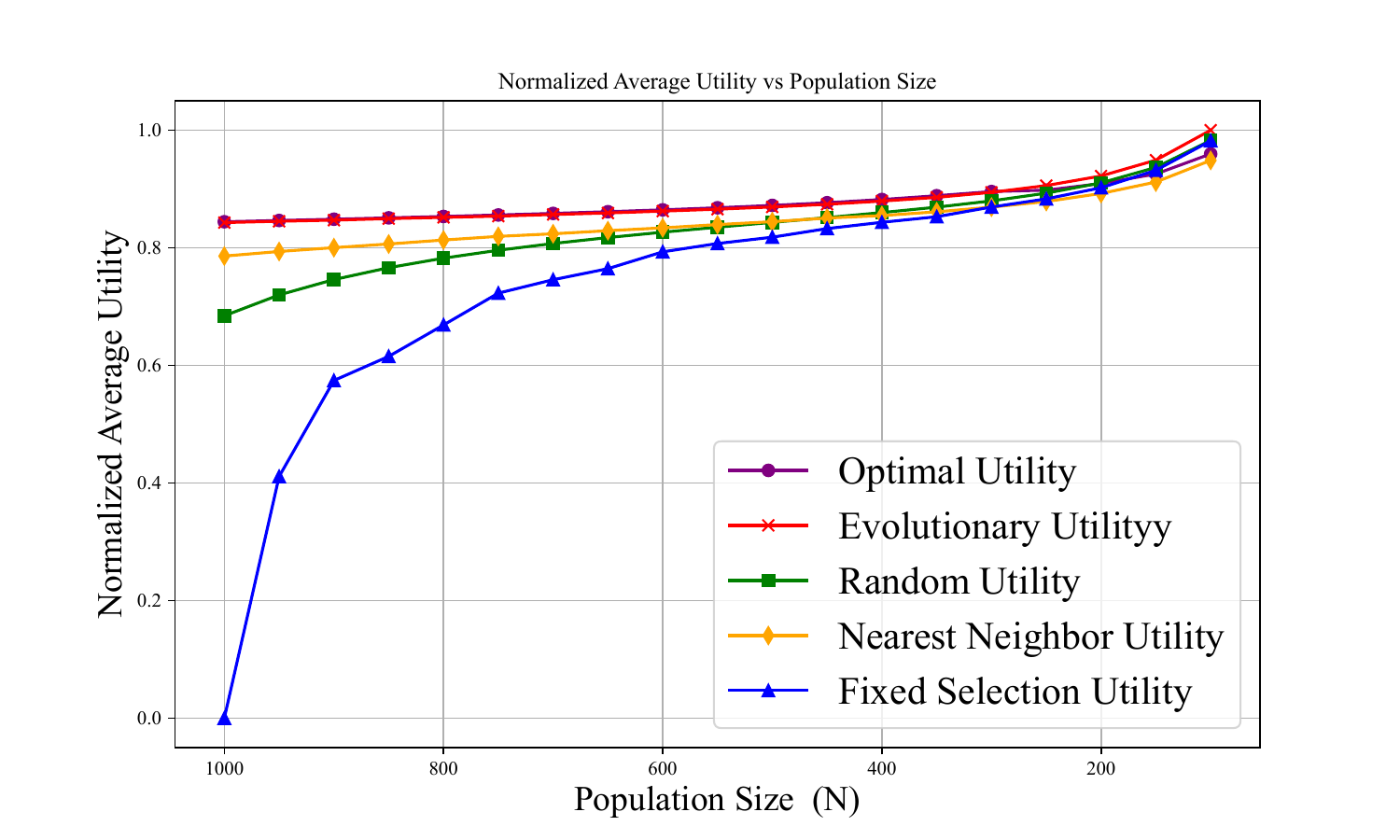}
			\caption{Normalized average utility with different population size.}
			\label{Fig:AverageUtility}
		\end{minipage}
		\hfill
		\begin{minipage}{0.3\textwidth}
			\centering
			\includegraphics[scale=0.25]{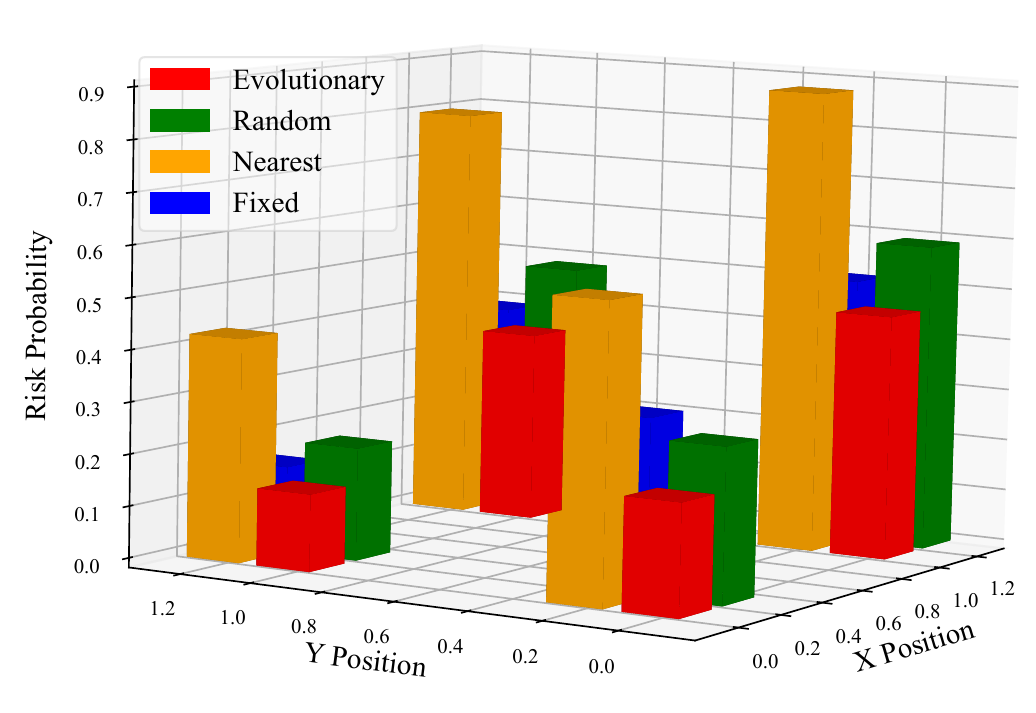}
			\caption{Risk probability comparison.}
			\label{Fig:RiskProbability}
		\end{minipage}
		\hfill
		\begin{minipage}{0.3\textwidth}
			\centering
			\includegraphics[scale=0.25]{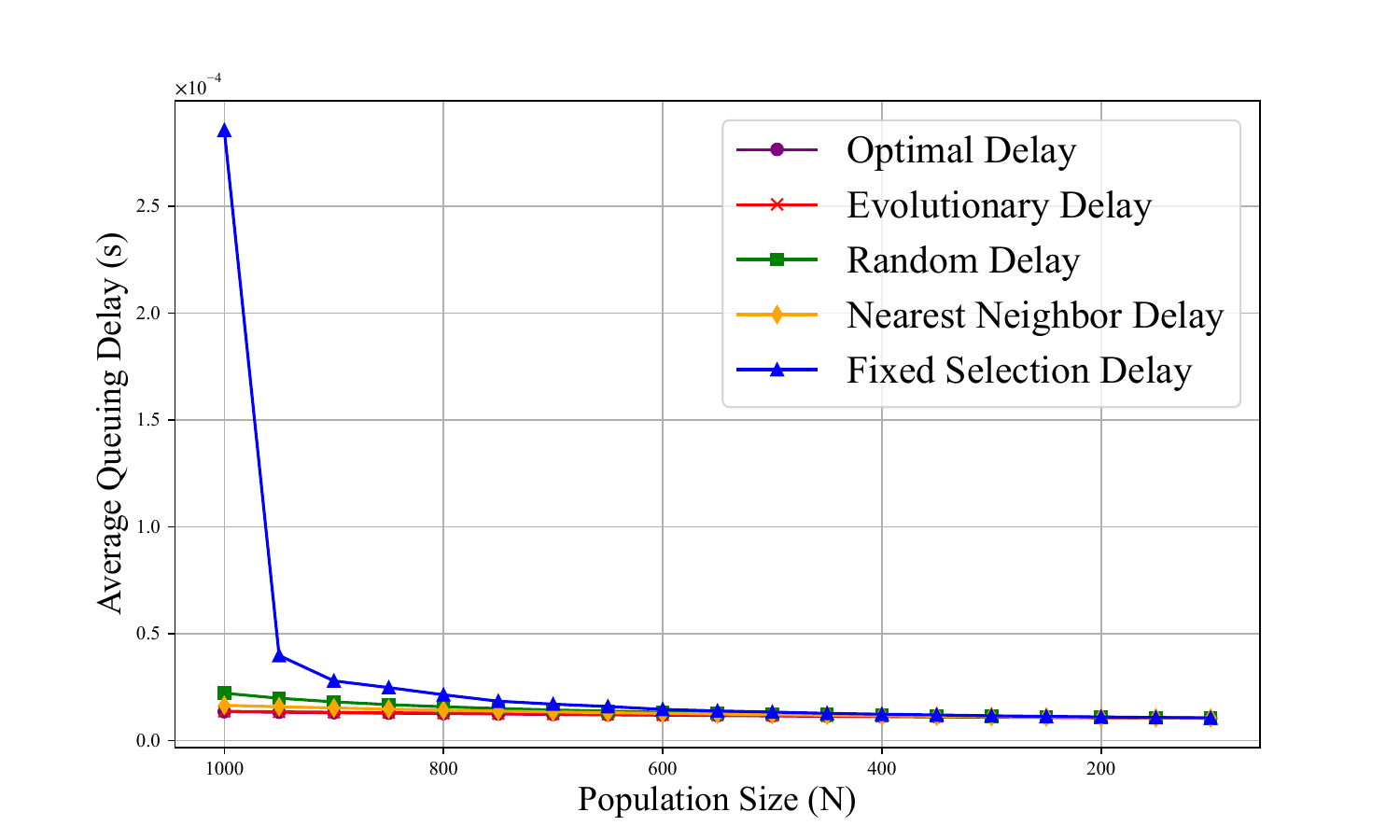}
			\caption{Queuing delay comparison.}
			\label{Fig:QueuingDelay}
		\end{minipage}
	\end{figure*}
	
	Fig. \ref{Fig:AverageUtility} compares the normalized average utility of five service selection strategies—optimal, evolutionary, random, nearest neighbor, and fixed selection—across different device population sizes. To eliminate scale-related deviations, we standardized the utility function parameters, ensuring that secrecy capacity and queuing delay are evaluated on the same scale, thus ensuring fair and accurate utility comparisons. Results show that the proposed method significantly outperforms other strategies in average utility and closely approaches the theoretical optimum. Additionally, the average utility of all methods increases as the number of devices decreases. This is because fewer devices lead to less competition for satellite resources, allowing more computing resources per device, reducing queuing delay, and improving resource allocation fairness. Thanks to its strong dynamic adjustment capability, the proposed algorithm adapts more effectively to changes in device density, achieving greater utility gains than other methods.

	\subsection{Risk Probability Comparison}

	To evaluate the security performance of various service selection strategies, this experiment quantifies the risk probability ${Pr}(t)$ that a satellite fails to meet the secrecy capacity demand of a device. ${Pr}(t)$ is defined and calculated using the formula:
	\equ{
		\operatorname{Pr}(t)= 
		\begin{cases}0 & \text {, if } SD_{n, i} \leq C_i^{\mathrm{S}}(t), \\ 
			1-e^{-\lambda_i\left(S D_{n, i}-C_i^{\mathrm{S}}(t)\right)} &\text {, if } SD_{n, i}>C_i^{\mathrm{S}}(t).
		\end{cases}
	}
	
	\noindent where, $SD_{n, i}$ represents the secrecy capacity demand of device $n$ for satellite $i$, while $C_i^{\mathrm{S}}(t)$ denotes the secrecy capacity provided by the satellite $i$. When the demand exceeds the satellite's capacity, the risk probability increases exponentially, highlighting the severity of the mismatch between the device demand and the satellite computing capacity.
	
	Fig. \ref{Fig:RiskProbability}  illustrates the risk probability for five service selection strategies across multiple scenarios, highlighting the impact of different strategies on security. Results show the evolutionary strategy reduces risk probability significantly through dynamic device selection, as it jointly optimizes service performance and security levels. In contrast, other non-adaptive methods (e.g., nearest selection) prioritize proximity over security, increasing vulnerability. The above analysis results further verify that it is of great significance to construct a service selection framework that reasonably integrates security indicators and balances the capabilities of devices and satellites.
	
	\subsection{Queuing Delay Comparison}

	The impact of five service selection strategies on the average queuing delay is evaluated. Fig. \ref{Fig:QueuingDelay} shows that the queuing delay for all strategies decreases as the population size declines. This phenomenon is due to the fact that a smaller population size means a lower competition pressure on satellite resources, allowing each device to be allocated more computing or communication resources, thereby reducing the overall queuing time. It is worth noting that among these strategies, the proposed selection strategy performs particularly well. The results are very close to the theoretical optimal solution. The balanced selection strategy significantly reduces the queuing delay by dynamically adjusting the relationship between devices and satellites. In contrast, other service selection strategies usually have higher queuing delays than the optimal and balanced selection. This difference arises because these strategies lack the ability to adjust access to satellites dynamically and do not consider satellite performance or load.
	
	\section{Conclusion}
	\label{Sec:Conclusion}
	
In this work, we propose a Digital Twin-driven secure access strategy for SAGIN-enabled IoT networks, addressing the challenges of offloading security-sensitive tasks under eavesdropping threats. Within the DT framework, potential risks are quantified through secrecy capacity modeling, enabling secure service selection. Leveraging an evolutionary game executed in the DT environment, a distributed algorithm dynamically balances secrecy capacity and queuing delay for adaptive satellite selection. Simulation results showed that the proposed strategy effectively reduced the probability of the system being subjected to eavesdropping risks and lowered queuing delay. Future research will focus on dynamic risk assessment, energy efficiency optimization, and cross-layer optimization for heterogeneous satellite networks.

	\ifCLASSOPTIONcaptionsoff
	\newpage
	\fi
	
	\bibliographystyle{IEEEtran}
	\bibliography{IEEEabrv,refer.bib}
	
\end{document}